\documentclass[preprint,prd,showpacs,amsmath,amssymb,amsthm,nofootinbib]{revtex4}
\usepackage[mathscr]{euscript}
\usepackage{bm}% bold math
\usepackage{graphicx}% Include figure files
\usepackage{subfigure}
\usepackage[colorlinks=true,linkcolor=red]{hyperref}
\newcommand{\bea}{\begin{eqnarray}}
\newcommand{\eea}{\end{eqnarray}}
\newcommand{\beq}{\begin{equation}}
\newcommand{\eeq}{\end{equation}}

\def\/{\over}

\begin{document}

\title{
Inflationary dynamics and preheating of the
non-minimally coupled inflaton field in the metric and
Palatini formalisms}

\author{ Chengjie Fu$^{1}$,  Puxun Wu$^{1,2}$\footnote{Corresponding author: pxwu@hunnu.edu.cn} and Hongwei Yu$^{1}$\footnote{Corresponding author: hwyu@hunnu.edu.cn}  }
\affiliation{$^1$Department of Physics and Synergetic Innovation Center for Quantum Effects and Applications, Hunan Normal University, Changsha, Hunan 410081, China \\
$^2$Center for High Energy Physics, Peking University, Beijing 100080, China }

\begin{abstract}

The inflationary dynamics and preheating in a model with a non-minimally coupled inflaton field in the metric and Palatini formalisms are studied in this paper.  We find that in both formalisms, irrespective of the initial conditions, our Universe will evolve into a slow-roll inflationary era  and then the scalar field  rolls into an oscillating phase. The value of the scalar field at the end  of the inflation in the Palatini formalism is always larger than that in the metric one, which becomes more and more obvious with the increase of  the absolute value of the coupling parameter $|\xi|$. During the preheating, we find that the inflaton quanta are produced explosively due to the parameter resonance and the growth of inflaton quanta will  be terminated by  the backreaction. With the increase of  $|\xi|$,  the resonance bands gradually  close to the zero momentum ($k=0$), and the  structure of resonance changes and becomes broader and broader in the metric formalism,  while, it remains to be narrow in the Palatini formalism.   The energy transfer  from the inflaton field to the fluctuation becomes more and more efficient with the increase of $|\xi|$, and  in the metric formalism  the growth of the efficiency of energy transfer is  much faster than that  in the Palatini formalism. Therefore,  the inflation and preheating  show different  characteristics in different formalisms.
\end{abstract}


\maketitle
%%%%%%%%%%%%%%%%%%%%%%%%%%%%%%%%%%%%%%%%%%%%%%%%%%
\section{Introduction}
\label{sec_in}
\setcounter{equation}{0}
%%%%%%%%%%%%%%%%%%%%%%%%%%%%%%%%%%%%%%%%%%%%%%%%%%
After inflation ends, the inflaton field rolls into the minimum of its potential and  oscillates around it.  During the oscillation the energy of the  inflaton field  transforms into some light particles and the Universe is reheated~\cite{EW}.  However, the preliminary theory of reheating fails to reach a temperature high enough to produce baryon asymmetry in  the grand unified theory  (GUT) when the reheating   ends~\cite{AD,LF}.   It is worthy of noting that in addition to the GUT baryogenesis, there are some other mechanisms, such as electroweak baryogenesis and  leptogenesis, which occur at much lower energy scales (see \cite{Dine} for recent review).  Later, it was realized that at the first stage of reheating there exists an explosive particle production  called preheating~\cite{JR,LAA,BSD}. In the simple chaotic inflation model~\cite{ADL} with the  potential $m^2\phi^2/2$, the preheating requires a new scalar field $\chi$ which is coupled to the inflaton field $\phi$ through  $g^2\phi^2\chi^2/2$~\cite{SI1,TT,LK}. The inflaton field rapidly decays to $\chi$-particle by the $g$-resonance. In the model with a self-coupling inflaton potential $\lambda\phi^4/4$, the fluctuation of the chaotic inflaton field was amplified by the parametric resonance and the inflaton field transforms into $\delta\phi$-particle \cite{SI2}. For the case of the $\lambda\phi^4/4+g^2\phi^2\chi^2/2$ potential , the above two processes  exist  simultaneously  and the value of the parameter $g^2/\lambda$ affects the structure of the resonance~\cite{PB}.

Although the usual chaotic inflation model with the  power-law potentials seems to have been disfavored by the observations from the Planck satellite~\cite{Planck} ,  a natural generalization of it that   assumes a coupling between the scalar field and the curvature is still consistent with the observations~\cite{Planck}.  The preheating process in the non-minimally coupled inflation models  in the metric formalism has been discussed in~\cite{SKT1, SKT2, DeCross}. When the inflaton potential is $m^2\phi^2/2$, it has been found that the non-minimal coupling enhances the $g$-resonance in certain parameter regimes~\cite{SKT1}. Once the potential is changed to  $\lambda\phi^4/4$, the non-minimal coupling modifies the structure of resonance and promotes the particle creation for the strong coupling~\cite{SKT2}. Actually, the coupling between the   kinetic term of the scalar field and the Einstein tensor is also allowed by observations~\cite{L, YTB}. However, because of the non-periodic oscillation of the inflaton field, there is no parametric resonance in this case~\cite{YY}.

 It is well known that,  in addition to the metric formalism, there is another choice called the Palatini formalism, where the metric tensor and the connection are both independent variables. In the Einstein theory of gravity, two different formalisms give the same field equations. Once the gravitational theory is modified, these two formalisms yield different field equations and lead to  different physical outcomes~\cite{TV,AS}. The inflation  in the Palatini formalism has been investigated in~ \cite{FD2,BDD,NC}, and it has been found that  scalar and tensor spectral indices and their ratio in different formalism show distinct results. In addition,   in the usual metric formalism    the non-minimally coupled Higgs inflation  faces the problem of the unitarity violation, and the whole inflationary evolution is dependent on the unknown ultraviolet (UV) physics \cite{CHM,RJ,Be11, Atk11, JJ}. While, once  the Palatini formalism is considered,  the Higgs inflation with the non-minimal coupling   is  UV-safe and  does not suffer from the problem which exists in the metric formalism~\cite{FD1}.  Furthermore, Rasanen and  Wahlman~\cite{Rasanen} found that   the   tensor-to-scalar ratio in  the non-minimally coupled Higgs inflation can be consistently suppressed  in the Palatini formalism compared to the one  in the metric formalism. So,  the non-minimally coupled scalar field inflation in the Palatini formalism  has been shown to exhibit interesting different features, even some advantages. 
  
  In this paper, we plan to carry out a further study on the inflationary dynamics with non-minimal couplings in both the  Palatini and metric formalisms, and investigate the properties of the preheating in these formalisms, i.e., the similarities and differences.
The remainder of this paper is organized as follows. Section \ref{sec_be} gives the basic field equations of the inflation model with non-minimal couplings. In Section \ref{sec_da}, we analyze the inflationary dynamics. We study the preheating process in Section \ref{sec_pr}. Finally, Section \ref{sec_co} presents the conclusions. Throughout this paper, unless specified, we adopt the metric signature ($-, +, +, +$). Latin indices run from 0 to 3 and the Einstein convention is assumed for repeated indices.

%%%%%%%%%%%%%%%%%%%%%%%%%%%%%%%%%%%%%%%%%%%%%%%%%%
\section{Basic equations}
\label{sec_be}
\setcounter{equation}{0}
%%%%%%%%%%%%%%%%%%%%%%%%%%%%%%%%%%%%%%%%%%%%%%%%%%

We consider an inflation model with a non-minimal coupling between the scalar field and   gravity described by the action:
\begin{equation}\label{1}
S=\int d^4x \sqrt{-g}\biggl[\frac{1}{2\kappa^2}F(\phi)\hat R+\mathcal{L}_\phi(g_{\mu\nu},\phi)\biggl]\;,
\end{equation}
where $\kappa^2/8\pi\equiv M^{-2}_{PL}$ with $M_{PL}$ being the Planck mass,  $g$ is the determinant of the metric tensor $g_{\mu\nu}$ and $\mathcal{L}_\phi$ is the Lagrangian of  scalar field $\phi$, which  has the form
\begin{equation}\label{2}
\mathcal{L}_\phi=-\frac{1}{2}\nabla_\mu\phi\nabla^\mu\phi-V(\phi)\;,
\end{equation}
with $ V(\phi)=\frac{1}{4}\lambda\phi^4$ being the self-interacting potential and $\lambda$ a constant. $\hat R$ is the curvature scalar, which, in the metric formalism,  is  determined only by the metric
tensor $g_{\mu\nu}$.  Hence, $\hat R=R(g_{\mu\nu})$. While, in the Palatini formalism, since the connection is also an independent variable, the curvature scalar $\hat R=\hat{R}(g_{\mu\nu},\hat{\Gamma}^\alpha_{\beta\gamma})$ is related to both the metric tensor $g_{\mu\nu}$ and the connection $\hat{\Gamma}^\alpha_{\beta\gamma}$. Appendix~\ref{app:gp} gives the basic equations of gravity in the Palatini formalism.

In general, $F$ is an arbitrary function of $\phi$. In this paper, we consider the following  special form:
\begin{eqnarray}\label{3}
F=1-\xi\kappa^2\phi^2\;.
\end{eqnarray}
where $\xi$ is the coupling constant, which is constrained to be $\xi\lesssim10^{-3}$~\cite{TK}. The tiny value of $\xi$  means an insignificant deviation from the minimal coupling if the coupling is positive. Thus, the  negative coupling  ($\xi \leq0$) will be considered in the following.

Varying the action~(\ref{1}) with respect to the metric tensor gives the field equations (see Appendix~\ref{app:gp} for detail):
\begin{align}\label{4}
G_{\mu\nu}=&\kappa^2\biggl[\nabla_\mu\phi\nabla_\nu\phi-g_{\mu\nu}\Bigl(\frac{1}{2}\nabla_\mu\phi\nabla^\mu\phi+V(\phi)\Bigl)\biggl]+(1-F)R_{\mu\nu}+\nabla_\mu\nabla_\nu F\notag \\
&+\frac{1}{2}g_{\mu\nu}\biggl[(F-1)R-2\nabla_\sigma\nabla^\sigma F\biggl]+\sigma\biggl[\frac{3}{4F}g_{\mu\nu}\nabla_\sigma F\nabla^\sigma F-\frac{3}{2F}\nabla_\mu F \nabla_\nu F\biggl]\;,
\end{align}
where $G_{\mu\nu}$  is the Einstein tensor, and $\sigma=1$ and $0$ correspond to the Palatini formalism and the metric one, respectively. The inflaton field satisfies the modified Klein-Gordon equation
\begin{equation}\label{5}
\square\phi-\xi\hat{R}\phi-V_{,\phi}=0\;,
\end{equation}
where $\square$ is defined as $\square\equiv\nabla_\mu\nabla^\mu$ and $V_{,\phi}=dV/d\phi$. From Eq.~(\ref{A8}), one can see that  the hatted curvature scalar $\hat R$ can be expressed as
\begin{equation}\label{6}
\hat{R}=R+\sigma\biggl[\frac{3}{2F^2}(\nabla_\mu F)(\nabla^\mu F)-\frac{3}{F}\nabla_\mu\nabla^\mu F\biggl]\;,
\end{equation}
or \begin{equation}\label{7}
\frac{F}{\kappa^2}\hat{R}=\nabla_\mu\phi\nabla^\mu\phi+4V+\frac{3(1-\sigma)}{\kappa^2}\nabla_\sigma\nabla^\sigma F\;.
\end{equation}
by using   Eqs.~(\ref{4}) and ~(\ref{6})  and $R=-g^{\mu\nu}G_{\mu\nu}$.

To investigate the amplification of the fluctuations of the inflaton field by the parameter resonance in the metric and Palatini formalisms,  we consider  the spatially flat Friedmann-Robertson-Walker metric:
\begin{equation}\label{8}
ds^2=-dt^2+a^2(t)d\mathbf{x}^2\;,
\end{equation}
where $t$ is the cosmic time and $a(t)$ is the scale factor. Since  the results  have  no significant difference  whether  the backreaction of metric perturbations is included~\cite{Zibin},     we only consider the  perturbations of the scalar field. Thus,  the inflaton field  $\phi$ consists of the homogeneous and fluctuant parts:
\begin{equation}\label{9}
\phi(t,\mathbf{x})=\phi_0(t)+\delta\phi(t,\mathbf{x})\;.
\end{equation}
For  the fluctuation, we impose the tadpole condition
\begin{equation}\label{10}
\langle\delta\phi(t,\mathbf{x})\rangle=0\;,
\end{equation}
where $\langle\cdot\cdot\cdot\rangle$ represents the expectation value, and   make use of the Hartree factorization:
\begin{align}
(\delta\phi)^3&\rightarrow 3\langle(\delta\phi)^2\rangle(\delta\phi)\;,\\
(\delta\phi)^4&\rightarrow 6\langle(\delta\phi)^2\rangle(\delta\phi)^2-3\langle(\delta\phi)^2\rangle^2\;.
\end{align}
Then the Friedmann equation with the backreaction of  the fluctuation is acquired via taking the mean field approximation in the $00$ component of Eq.~(\ref{4}):
\begin{align}\label{13}
H^2&=\frac{\kappa^2}{3(1-\alpha)}\biggl[\frac{1}{2}(\dot{\phi}_0^2+\langle\delta\dot{\phi}^2\rangle)+\bigg(\frac{1}{2}-2\xi \bigg )\langle\delta\phi^{\prime2}\rangle+\frac{1}{4}(\phi_0^4+6\phi_0^2\langle\delta
\phi^2\rangle+3\langle\delta\phi^2\rangle^2)\nonumber\\ &+2\xi\{3H(\phi_0\dot\phi_0+\langle\delta\phi\delta\dot\phi\rangle)-\langle\delta\phi\delta\phi^{\prime\prime}\rangle\}-\frac{3\sigma\xi^2\kappa^2}{1-\alpha} \big (\phi_0^2\dot\phi_0^2+4\phi_0\dot\phi_0\langle
\delta\phi\delta\dot\phi\rangle.  \nonumber\\
&+\dot\phi_0^2\langle\delta\phi_0^2\rangle+\phi_0^2\langle\delta\dot\phi^2\rangle+\langle\delta\phi^2\delta\dot\phi^2\rangle+
\phi_0^2\langle\delta\phi^{\prime2}\rangle+\langle\delta\phi^2\delta\phi^{\prime2}\rangle \big )\biggl]\;,
\end{align}
where $\alpha\equiv\xi\kappa^2\langle\phi^2\rangle$, $H=\dot{a}/a$ is the Hubble parameter,  and a dot and a prime denote the derivative with respect to the cosmic time and the space coordinate, respectively.

Taking the mean field approximation of Eq.~(\ref{5}), we  obtain that   the inflation field and its perturbation  obey
\begin{equation}\label{15}
\ddot\phi_0+3H\dot\phi_0+\lambda\phi_0(\phi_0^2+3\langle\delta\phi^2\rangle)+\xi \hat R \phi_0=0\;,
\end{equation}
\begin{equation}\label{16}
\delta\ddot\phi+3H\delta\dot\phi-\partial_i\partial^i(\delta\phi)+\{3\lambda(\phi_0^2+\langle\delta\phi^2\rangle)+\xi \hat R\}\delta\phi=0\;,
\end{equation}
respectively,  where the hatted curvature scalar $\hat R$ has the form
\begin{align}\label{14}
\hat R&=\frac{\kappa^2}{1-\alpha}\bigl[(-\dot\phi_0^2-\langle\delta\dot\phi^2\rangle+\langle\delta\phi^{\prime2}\rangle)+\lambda(\phi_0^4+6\phi_0^2\langle\delta\phi^2\rangle+3\langle\delta\phi^2\rangle)
\nonumber\\&
+(1-\sigma)\{-6\xi(-\dot\phi_0^2-\langle\delta\dot\phi\rangle+\langle\delta\phi^{\prime2}\rangle)+6\xi(\phi_0\ddot\phi_0+\langle\delta\phi\delta\ddot\phi\rangle\nonumber\\&
-\langle\delta\phi\delta\phi^{\prime\prime}\rangle)+18\xi H(\phi_0\dot\phi_0+\langle\delta\phi\delta\dot\phi\rangle)\}\bigl]\;.
\end{align}
Since the fluctuant part of the inflation field can be expanded as
\begin{equation}\label{17}
\delta\phi=\frac{1}{(2\pi)^{3/2}}\int (a_k\delta\phi_k(t)e^{-i\mathbf{k}\cdot\mathbf{x}}+a_k^{\dagger}\delta\phi_k^{\ast}(t)e^{i\mathbf{k}\cdot\mathbf{x}})d^3\mathbf{k}\;,
\end{equation}
where $a_k$ and $a_k^\dagger$ are the annihilation and creation operators respectively and $k=|\mathbf{k}|$, from Eq.~(\ref{16}), we find that $\delta\phi_k$ satisfies the following equation of motion:
\begin{equation}\label{18}
\delta\ddot\phi_k+3H\delta\dot\phi_k+\biggl[\frac{k^2}{a^2}+3\lambda(\phi_0^2+\langle\delta\phi^2\rangle)+\xi \hat R\biggl]\delta\phi=0
\end{equation}
with the expectation values of $\delta\phi^2$  taking the form
\begin{equation}\label{}
\langle\delta\phi^2\rangle=\frac{1}{2\pi^2}\int k^2|\delta\phi_k|^2dk\;.
\end{equation}

Introducing the conformal time $\eta\equiv\int a^{-1}dt$ and defining a new scalar field $\varphi_0\equiv a\phi_0$ and its perturbation $\delta\varphi \equiv a \delta\phi $, Eqs.~(\ref{15}) and~(\ref{18}) can be rewritten as
\begin{equation}\label{21}
\frac{d^2\varphi_0}{d\eta^2}+\lambda\varphi_0(\varphi_0^2+3\langle\delta\varphi^2\rangle)+(\xi \hat R a^2-\frac{1}{a}\frac{d^2a}{d\eta^2})\varphi_0=0\;,
\end{equation}
\begin{equation}\label{22}
\frac{d^2}{d\eta^2}\delta\varphi_k+\omega_k^2\delta\varphi_k=0\;,
\end{equation}
 with
\begin{equation}\label{23}
\omega_k^2\equiv k^2+3\lambda(\varphi^2_0+\langle\delta\varphi^2\rangle)+\xi \hat R a^2-\frac{1}{a}\frac{d^2a}{d\eta^2}\;.
\end{equation}
In the non-minimal coupling model, the $\xi \hat R a^2$ term affects not only the oscillation mode of the inflaton field but also the parametric resonance. Furthermore, the contributions of this term in the metric formalism are different  from those in the Palatini formalism.

We choose the conformal vacuum as the initial state and the following initial conditions of the fluctuation
\begin{align}
\delta\varphi_k(0)&=\frac{1}{\sqrt{2\omega_k(0)}}\;,\\
\delta\dot\varphi_k(0)&=-i\omega_k(0)\delta\varphi_k(0)\;.
\end{align}
Before investigating the preheating process, we will study the evolution of the scale factor in the inflationary period and analyze the dynamical behavior of the inflaton field.

%%%%%%%%%%%%%%%%%%%%%%%%%%%%%%%%%%%%%%%%%%%%%%%%%%
\section{The  inflationary dynamics}
\label{sec_da}
%\setcounter{equation}{0}
%%%%%%%%%%%%%%%%%%%%%%%%%%%%%%%%%%%%%%%%%%%%%%%%%%

In this Section, we investigate the dynamical property of  the inflaton field with a non-minimal coupling. In the inflationary period, the fluctuation of the inflaton field can be neglected. Then,
Eqs.~(\ref{13}) and~(\ref{15}) are simplified as
\begin{equation}\label{26}
H^2=\frac{\kappa^2}{3(1-\alpha_0)}\biggl[\frac{1}{2}\dot\phi_0^2+\frac{1}{4}\lambda\phi^4_0+6\xi H\phi_0\dot\phi_0-\frac{3\sigma\xi^2\kappa^2\phi_0^2\dot\phi_0^2}{1-\alpha_0}\biggl]\;,
\end{equation}
\begin{equation}\label{29}
\ddot\phi_0+3H\dot\phi_0+\frac{\lambda\phi_0^3}{1-(1-(1-\sigma)6\xi)\alpha_0}-\frac{\xi\kappa^2(1-(1-\sigma)6\xi)\phi_0\dot\phi_0^2}{1-(1-(1-\sigma)6\xi)\alpha_0}=0\;,
\end{equation}
where $\alpha_0=\xi\kappa^2\phi_0^2$ and Eq.~(\ref{14}) has been used in obtaining Eq.~(\ref{29}).

To obtain the inflation, one needs to  define the slow-roll conditions: $|\ddot\phi_0/\dot\phi_0| \ll H$, $|\dot\phi_0/\phi_0| \ll H$ and $\dot\phi_0^2\ll V$. Meanwhile, to see the effect of the non-minimal coupling, we take the strong negative
 coupling limit $\alpha_0\ll-1$. Thus, in the inflationary era, Eqs.~(\ref{26}) and (\ref{29}) reduce to 
\begin{equation}\label{30}
H^2\simeq-\frac{\kappa^2\lambda\phi_0^4}{12\alpha_0}\;,
\end{equation}
\begin{equation}\label{31}
3H\dot\phi_0\simeq\frac{\lambda\phi_0^3}{(1-(1-\sigma)6\xi)\alpha_0}\;.
\end{equation}
Using Eqs. (\ref{30}) and (\ref{31}), we obtain the expression of the e-fold number:
\begin{align}\label{32}
N&\equiv\int_{t_I}^{t_F}Hdt=\int_{\phi_0(t_I)}^{\phi_0(t_F)}\frac{H}{\dot\phi_0}d\phi_0=-\int_{\phi_0(t_I)}^{\phi_0(t_F)}\frac{(1-(1-\sigma)6\xi)\kappa^2}{4}\phi_0d\phi_0\nonumber\\ &=\frac{(1-(1-\sigma)6\xi)\kappa^2}{8}[ \phi_0(t_I)^2-\phi_0(t_F)^2],
\end{align}
where the subscript $I$ and $F$ denote the values when the inflation starts and ends, respectively. Considering $\phi_0(t_F)\ll\phi_0(t_I)$, we arrive at  the initial value of $\phi_0$ field  
\begin{equation}\label{33}
\phi_0(t_I)^2=\frac{8N}{(1-(1-\sigma)6\xi)\kappa^2}.
\end{equation}
In addition, combining Eqs. (\ref{30}) and (\ref{31}), we find the following relation:
\begin{equation}\label{34}
\dot\alpha_0=-\frac{8\xi}{1-(1-\sigma)6\xi} H\;,
\end{equation}
which is easily integrated to give
\begin{equation}\label{35}
\alpha_0=\alpha_{0I}-\frac{8\xi}{1-(1-\sigma)6\xi}\ln\left(\frac{a}{a_I}\right)\;.
\end{equation}
From Eq. (\ref{30}), it follows that the scale factor satisfies the  differential equation:
\begin{equation}\label{36}
\frac{\dot a}{a}=\sqrt{\frac{-\lambda\alpha_{0I}}{12\xi^2\kappa^2}}\biggl[1-\frac{8\xi}{(1-(1-\sigma)6\xi)\alpha_{0I}}\ln\left({\frac{a}{a_I}}\right)\biggl]^{1/2}\;.
\end{equation}
Integrating the above equation gives
\begin{equation}\label{37}
a=a_I \exp[H_I t-\gamma(H_I t)^2]\;,
\end{equation}
where
\begin{equation}\label{38}
H_I\equiv \sqrt{\frac{-\lambda\alpha_{0I}}{12\xi^2\kappa^2}}=\sqrt{\frac{-2\lambda N}{3(1-(1-\sigma)6\xi)\xi\kappa^2}}\;,\qquad\gamma\equiv\frac{2\xi}{(1-(1-\sigma)6\xi)\alpha_{0I}}=\frac{1}{4N}\;.
\end{equation}
Apparently $\gamma$ is a small quantity since $N\gtrsim60$. At the beginning of inflation, the Universe evolves exponentially since the $\gamma$ term is negligible. However, as time goes on, this $\gamma$ term becomes significant and the expansion rate of the universe slows down. The value of $\gamma$ is the same in different formalisms, but the value of $H_I$ in the Palatini formalism is about $\sqrt{-6\xi}$ times of that in the metric formalism, which means that in the  strong  coupling  case the deceleration effect caused by the $\gamma$ term will be greater in the Palatini formalism than  in the metric formalism.

\begin{table}
\caption{\label{Tab1}The  values of $\phi_0$ when the inflation ends in the metric and Palatini formalisms for different $\xi$. }
\begin{center}
  \begin{tabular}{c|cc}
  \hline
  \hline
  &$\qquad\qquad\qquad\qquad\qquad\phi_0(t_F)/M_{PL}$\\
  \hline
  $\xi$$\qquad$ &Palatini $\qquad$& $\qquad$Metric$\qquad$\\
  \hline
  0$\qquad$ & 0.5642 $\qquad$ & $\qquad$ 0.5642$\qquad$\\
  -0.1$\qquad$ & 0.4569 $\qquad$ & $\qquad$ 0.4281$\qquad$\\
  -1$\qquad$ & 0.3072 $\qquad$ &  $\qquad$ 0.1995$\qquad$\\
  -10$\qquad$ & 0.1835 $\qquad$ &  $\qquad$ 0.0673$\qquad$\\
  -50$\qquad$ & 0.1246 $\qquad$ &   $\qquad$ 0.0303$\qquad$\\
  -200$\qquad$ & 0.0887 $\qquad$ &  $\qquad$ 0.0152$\qquad$\\
  -1000$\qquad$ & 0.0595 $\qquad$ &  $\qquad$ 0.0068$\qquad$\\
  \hline
  \hline
  \end{tabular}
\end{center}
\end{table}

Now we want to determine the value of $\phi_0(t_F)$, which corresponds to the initial value of reheating. For convenience, we change to the Einstein frame. The detailed calculation can be found in   Appendix \ref{app:ae}. In the Einstein frame  one can define the slow-roll parameter by using the potential $\bar V(\psi)$ given in Eq.~(\ref{B3})\begin{equation}\label{39}
\epsilon=\frac{1}{\kappa^2}\biggl(\frac{\bar d V(\psi)/d\psi}{\bar V(\psi)}\biggl)^2\;.
\end{equation}
When $\epsilon$ increases to unity,  the inflation ends. Using Eqs.~(\ref{B3})-(\ref{B5}), the slow-roll parameter can be rewritten as
\begin{equation}\label{40}
\epsilon=\frac{8}{\kappa^2\phi_0^2[1-(1-(1-\sigma)6\xi)\xi\kappa^2\phi_0^2]}\;.
\end{equation}
Setting $\epsilon=1$, we obtain the value of the scalar field $\phi_0$ at the end of the inflation:
\begin{equation}\label{41}
\phi_0(t_F)=\biggl[\frac{\sqrt{1-32\xi+(1-\sigma)192\xi^2 }-1}{16\pi(1-(1-\sigma)6\xi)|\xi|}\biggl]^{1/2}M_{PL}\;.
\end{equation}
Tab.~\ref{Tab1} shows the values of $\phi_0(t_F)$ with different values of $\xi$ in the metric and Palatini formalisms. With the increase of $|\xi|$, $\phi_0(t_F)$ decreases in both formalisms. But for the same value of the  coupling parameter,    $\phi_0(t_F)$  in the Palatini formalism is always larger than that in the metric one, which becomes more and more obvious with the increase of $|\xi|$.

In order to show clearly the  characteristics  of the non-minimally coupled inflation, we turn to the dynamics which is  governed by Eqs.~(\ref{26}, \ref{29}).
A general result for an inflationary attractor is very hard to  obtain. But, in the strong coupling limit: $\xi\ll-1$,   Eqs. (\ref{30}) and (\ref{31}) give rise to  the inflationary attractors of the form
\begin{align}\label{48}
\dot\phi_0\simeq\left\{\begin{array}{lcl}
-\frac{1}{\kappa^2(1-(1-\sigma)6\xi)}\sqrt{\frac{4\lambda}{-3\xi}}\;,\quad(\phi_0>0)\\
+\frac{1}{\kappa^2(1-(1-\sigma)6\xi)}\sqrt{\frac{4\lambda}{-3\xi}}\;,\quad(\phi_0<0)\end{array}\right.\;.
\end{align}

\begin{figure}[!htb]
                \centering
                             \subfigure[$\sigma=1$ (Palatini)]{
                \includegraphics[width=0.47\textwidth ]{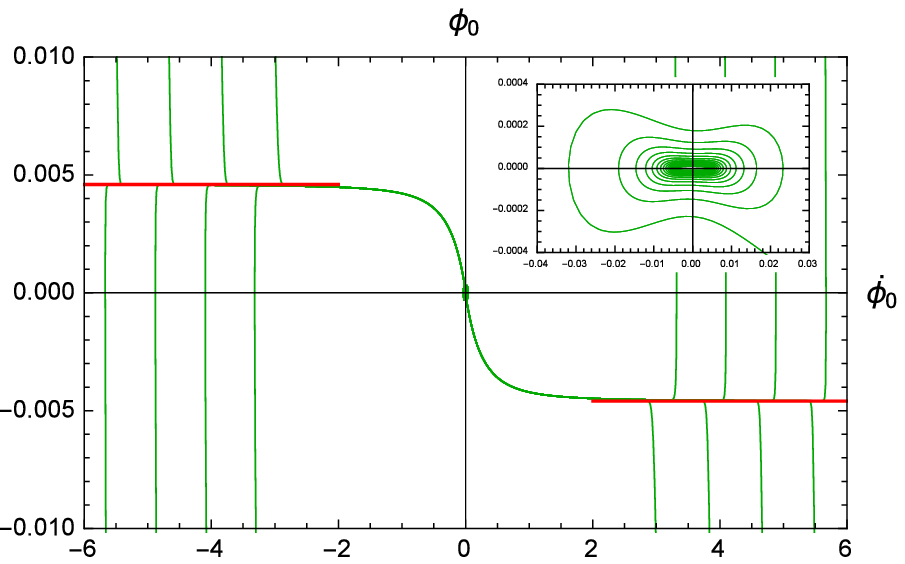}}
                   \subfigure[$\sigma=0$ (Metric)]{
                 \includegraphics[width=0.49\textwidth ]{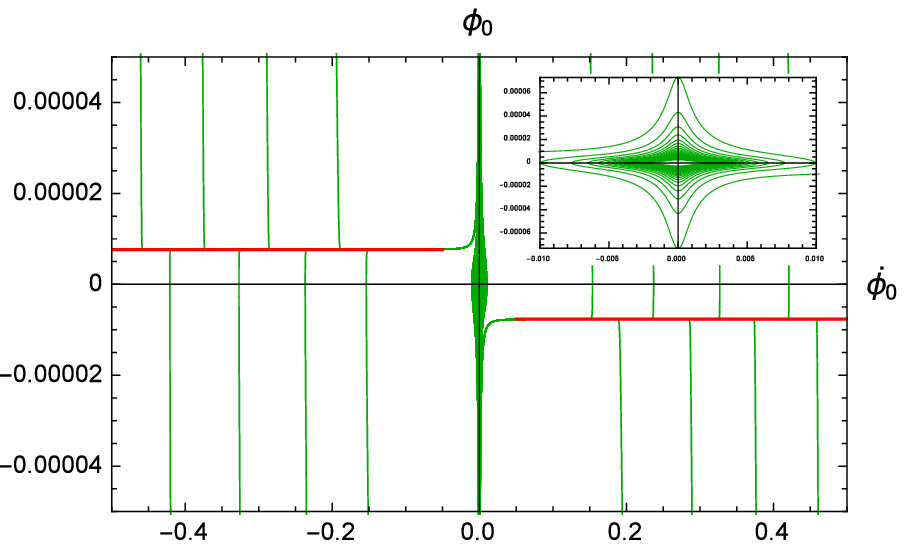}}
                   \caption{\label{Fig1} The phase portrait in the $\phi_0-\dot\phi_0$ plane with  $\xi=-100$,  $\lambda=1$ and $M_{PL}=1$. The red lines  describe the inflationary attractors given in Eq. (\ref{48}).}
\end{figure}

Fig. \ref{Fig1} shows the evolutionary curves in the $\phi_0-\dot{\phi}_0$ plane with different initial conditions.  The left and right panels denote the Palatini and metric formalisms, respectively. The red lines represent the inflationary attractors. It is easy to see that regardless of the initial conditions the Universe dominated by a non-minimally coupled scalar field will evolve into an inflation and then go into an oscillating era, which represents the reheating phase.  Comparing the left and right  panels  shows the apparent differences between the metric and the Palatini formalisms.
One can find that the evolution of the inflaton field in the Palatini formalism deviates from the slow-roll line with a value of $\phi_0$ much larger than that in the metric formalism. This is because the former has a  value of $H_I$ as given in Eq.~(\ref{38})   larger than the latter.  After the inflation,   the inflaton field rolls down along the potential with the increasing velocity in the metric formalism  but the velocity  is decreasing in the Palatini formalism. After the inflaton field rolls through the point of $\phi_0=0$, the value of $\dot\phi$ increases in the Palatini formalism, while it decreases in the metric formalism. When the scalar field evolves through its minimum and maximum,  the change of the velocity of the inflaton field in the metric formalism is more  violent than that in the Palatini formalism.  Therefore, the oscillation of the inflaton filed in different formalisms shows distinctive features, which leads to  a different preheating process as will be studied in the next section.

%%%%%%%%%%%%%%%%%%%%%%%%%%%%%%%%%%%%%%%%%%%%%%%%%%
\section{Preheating}
\label{sec_pr}
%\setcounter{equation}{0}
%%%%%%%%%%%%%%%%%%%%%%%%%%%%%%%%%%%%%%%%%%%%%%%%%%
From the above section, one can see that after the inflation the scalar field evolves into an oscillating era, in which the energy of  the scalar field transforms  into some light particles and our Universe is reheated. The first stage of reheating is called  preheating, which corresponds  to the generation of particles due to the parameter resonance.

Setting  $t_i$ to be the beginning time of preheating,  one can assume  reasonably  $t_i=t_F$ with $t_F$ being the finishing time of inflation, so that  the  value of the scalar field at the beginning of  preheating  is $\phi_0(t_i)=\phi_0(t_F)$, where $\phi_0(t_F)$ is given in Eq.~(\ref{41}).  After  introducing a dimensionless time variable $x\equiv\sqrt{\lambda}M_{PL}\eta$ and defining  $f\equiv\varphi_0/M_{PL}=a\phi_0/M_{PL}$, $\langle\delta\bar\phi^2\rangle\equiv\langle\delta\phi^2\rangle/M_{PL}^2$ and
$\langle\delta\bar\varphi^2\rangle\equiv a^2\langle\delta\bar\phi^2\rangle$, Eqs.~(\ref{21}) and (\ref{22}) can be re-expressed as
\begin{equation}\label{50}
\frac{d^2f}{dx^2}+f^3\left(1+3\frac{\langle\delta\bar\varphi^2\rangle}{f^2}\right)+\left(\xi a^2 \mathcal{\hat R}-\frac{1}{a}\frac{d^2a}{dx^2}\right)f=0\;,
\end{equation}
\begin{equation}\label{51}
\frac{d^2}{dx^2}\delta\varphi_k+\biggl[p^2+3f^2\bigg(1+\frac{\langle\delta\bar\varphi^2\rangle}{f^2}\bigg)+\bigg(\xi a^2\mathcal{\hat R}-\frac{1}{a}\frac{d^2a}{dx^2}\bigg)\biggl]\delta\varphi_k=0\;,
\end{equation}
where $p^2\equiv k^2/(\lambda M_{PL}^2)$, and $\mathcal{\hat R}\equiv \hat R/(\lambda M_{PL}^2)$, which has the form
\begin{equation}\label{52}
\mathcal{\hat R}\approx\frac{8\pi (1-(1-\sigma)6\xi)( \bar\phi_0^4-a^{-2}(d\bar\phi_0/dx)^2)}{1-(1-(1-\sigma)6\xi)\alpha}
\end{equation}
at the stage before the inflaton quanta  grow significantly. Here $\bar\phi_0\equiv\phi_0/M_{PL}$ and   Eqs.~(\ref{29}) has been used.  In order to analyze the effect of the non-minimal coupling on the growth  of inflaton quanta,   we introduce two frequencies, according to Eq.~(\ref{51}), 
  \begin{eqnarray}
  \omega_1\equiv3f^2(1+\langle\delta\bar\varphi^2\rangle/f^2), \qquad \omega_2\equiv  \xi a^2\mathcal{\hat R}-\frac{1}{a}\frac{d^2a}{dx^2} \;.
  \end{eqnarray}
  For the case of a very weak coupling, i.e. $-0.1\lesssim\xi<0$,  the effect of the non-minimal coupling is negligible. The evolutions of $\omega_1$ and $\omega_2$ in both formalisms are almost the same as what were obtained in the minimally coupled case~\cite{PB,SKT2}.

When the coupling becomes very strong, i.e. $\xi\lesssim-100$,  we can take the limit of strong coupling.   Assuming that  at the initial stage of preheating  the amplitude of the oscillating scalar field is not very small, we have $\xi \gg \xi \alpha \gg 1$, which means that   Eq.~(\ref{52}) can be simplified  to be
\begin{eqnarray}
\mathcal{R}\simeq \frac{8\pi}{\alpha}\big(a^{-2}(d \bar{\phi}_0/dx)^2- \bar{\phi}_0^4 \big)
\end{eqnarray}
in the metric formalism and
\begin{eqnarray}
\mathcal{\hat{R}}\simeq \frac{8\pi}{1-\alpha}\big(\bar{\phi}_0^4-a^{-2}(d \bar{\phi}_0/dx)^2\big)\simeq - {\alpha}
\mathcal{R}
\end{eqnarray}
in the Palatini formalism. Then, using Eq.~(\ref{52}) one can obtain
\begin{eqnarray}
\omega_2=\bigg(\xi-\frac{1}{6} \bigg) a^2 \mathcal{ R} \simeq \xi a^2 \mathcal{ R} \end{eqnarray}
in the metric formalism and 
\begin{eqnarray}
\omega_2=\xi a^2 \mathcal{\hat{R}}-\frac{1}{6} a^2 \mathcal{R} \simeq -\xi \alpha a^2 \mathcal{ R}  \end{eqnarray}
in the Palatini formalism.
Thus, in the metric formalism the perturbation term $\frac{1}{a}\frac{d^2a}{dx^2}$ is magnified $6|\xi|$ times,  suggesting that the decrease of   $\omega_2$  is lower than $\omega_1$ with the increase of $|\xi|$.  
In the Palatini formalism, the $\xi a^2 \mathcal{\hat R}$ term is also amplified with the increase of $|\xi|$ but it is much less than that in the metric formalism.  

\begin{table}
\caption{\label{Tab2}The  maximal values  $\langle \delta \bar \phi^2\rangle_f$, the oscillation amplitude of the inflaton field $\tilde{\phi}_0(x_f)$, and the ratio of them ($\langle \delta \bar\phi^2\rangle_f/  \tilde{\phi}^2_0(x_f)$) in the metric and Palatini formalisms for different $\xi$. }
\begin{center}
  \begin{tabular}{c|cc|cc|cc}
  \hline
  \hline
  &$\qquad \langle \delta \bar \phi^2\rangle_f/M^2_{PL}$& &$\qquad  \tilde{\phi}_0(x_f)/M_{PL}$ & &$ \qquad  \langle \delta \bar\phi^2\rangle_f/  \tilde{\phi}^2_0(x_f) $\\
  \hline
  $\xi$ &Palatini  & Metric & Palatini  & Metric  & Palatini  & Metric \\
  \hline
  0  & $1.460\times10^{-7}$  &   $1.460\times10^{-7}$  & $1.35\times10^{-3}$  &   $1.35\times10^{-3}$ & 0.08 &0.08\\
  -1  & $1.454\times10^{-7}$   &   $1.451\times10^{-7}$ & $1.30\times10^{-3}$   &   $1.25\times10^{-3}$  & 0.09 &0.09\\
%  -20  &$1.445\times10^{-7}$   &   $9.619\times10^{-8}$ & $1.14\times10^{-3}$   &   $6.24\times10^{-4}$  & 0.11 &0.24\\
  -50  & $1.403\times10^{-7}$   &   $7.642\times10^{-8}$ & $1.04\times10^{-3}$   &   $3.67\times10^{-4}$  &0.13 & 0.56\\
%  -100& $1.415\times10^{-7}$   &   $1.548\times10^{-7}$ & $ {\bf 1.04\times10^{-3}}$   &   $ \bf3.67\times10^{-4}$  &  \bf 0.13 & \bf 0.56\\
  -200 &$1.414\times10^{-7}$   &   $1.709\times10^{-6}$ & $8.56\times10^{-4}$   &   $9.50\times10^{-4}$  & 0.19 &1.88\\
  \hline
  \hline
  \end{tabular}
\end{center}
\end{table}

\begin{figure}[!htb]
                \centering
\subfigure[ Palatini]{                \includegraphics[width=0.45\textwidth ]{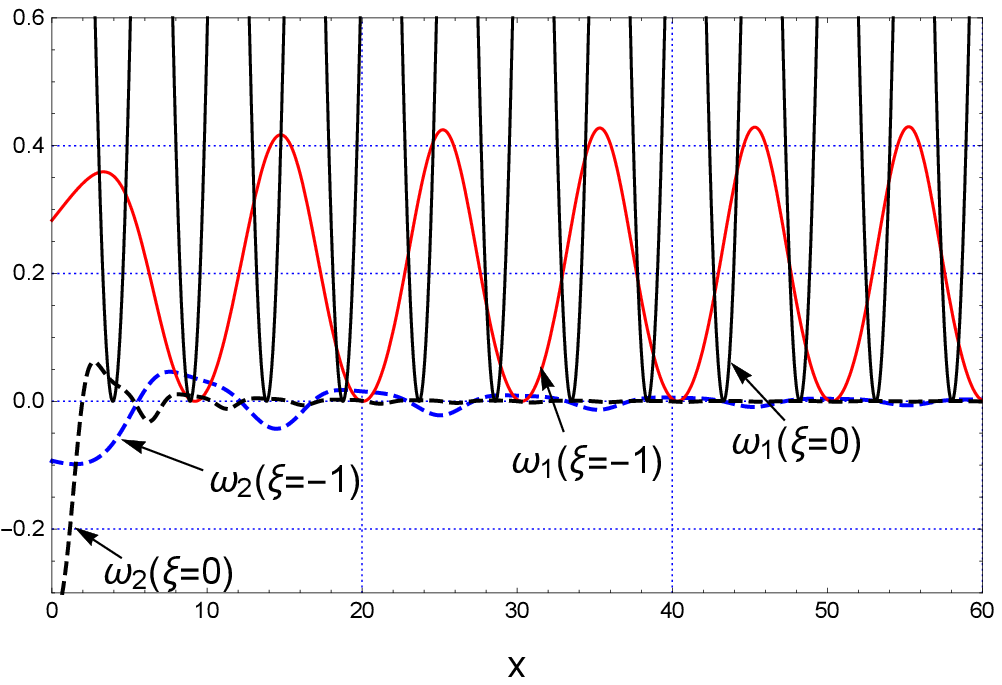}}
\subfigure[ Metric]{                 \includegraphics[width=0.45\textwidth ]{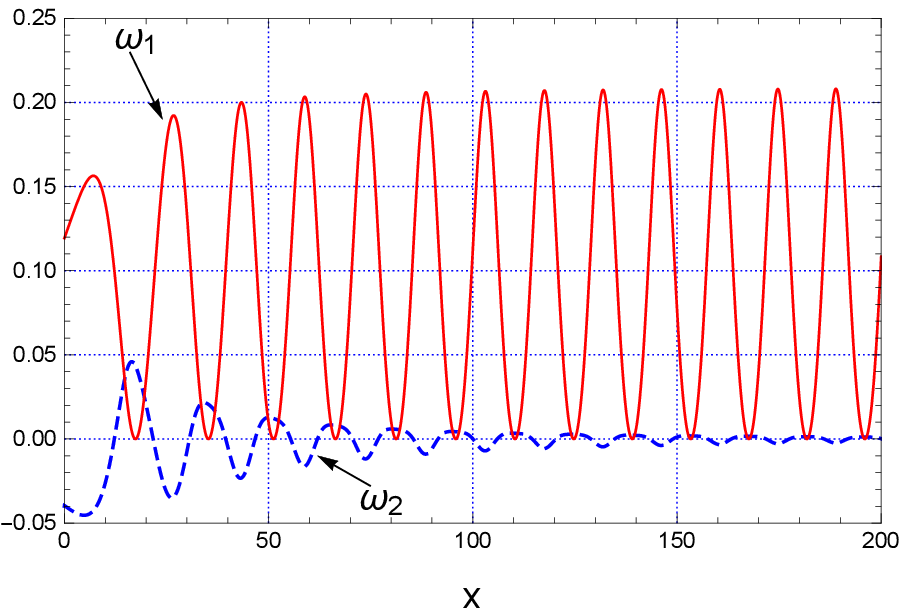}}
                \caption{\label{Fig2} The evolution of $\omega_1$ and $\omega_2$ as a function of $x$ in the case of $\xi=-1$. For a comparison, the evolution in the minimally coupled case is shown with the black curves in the left panel.}
\end{figure}
\begin{figure}
\subfigure[ Palatini]{
\includegraphics[width=0.45\textwidth]{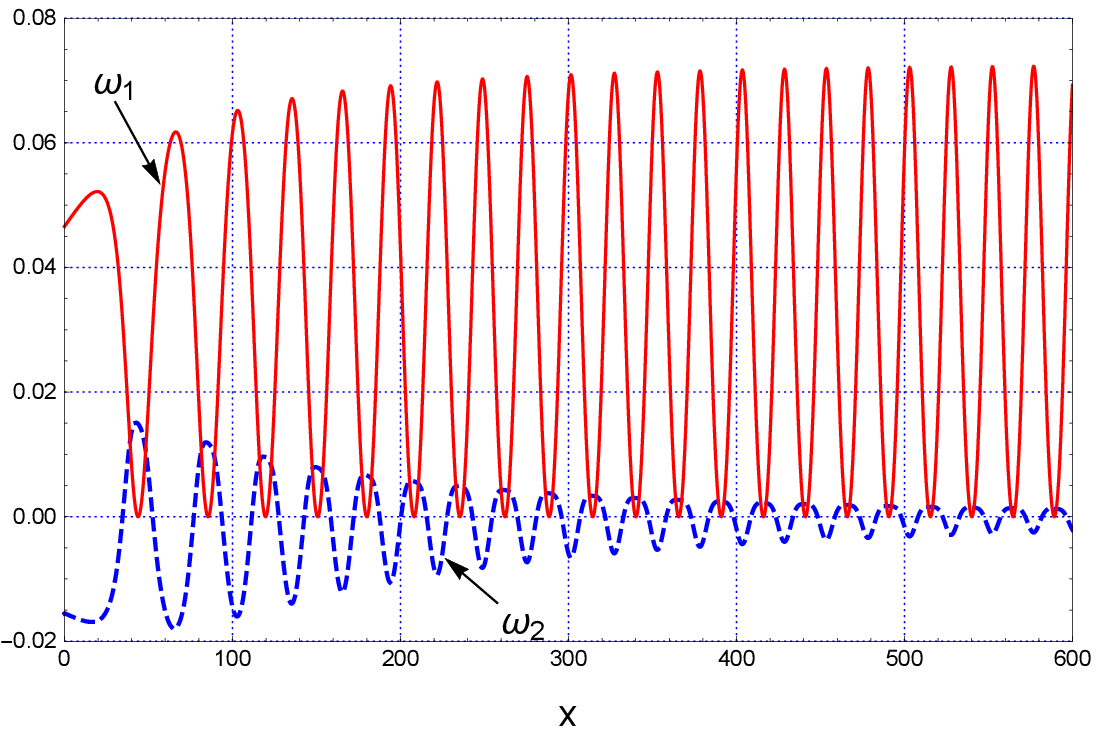}
}
\subfigure[ Metric]{
\includegraphics[width=0.45\textwidth]{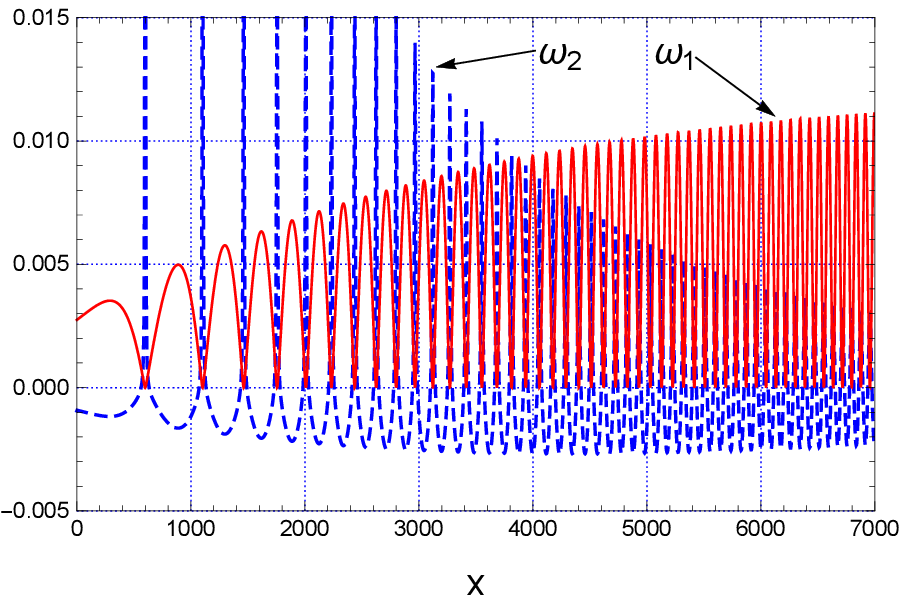}
}
\caption{\label{Fig3} The evolution of $\omega_1$ and $\omega_2$ as a function of $x$ in the case of $\xi=-50$.  }
\end{figure}

\begin{figure}
\subfigure[Palatini]{
\includegraphics[width=0.45\textwidth]{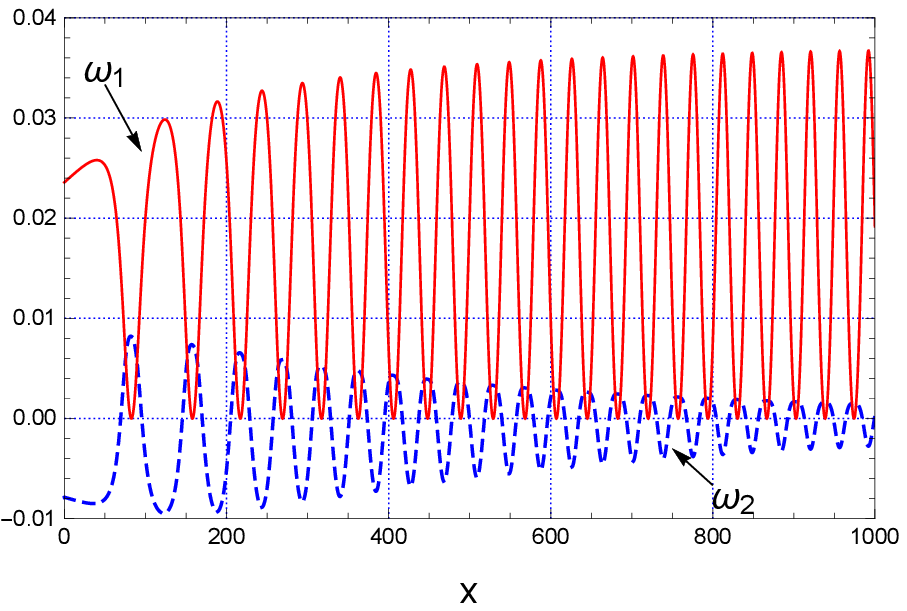}}
\subfigure[Metric]{
\includegraphics[width=0.45\textwidth]{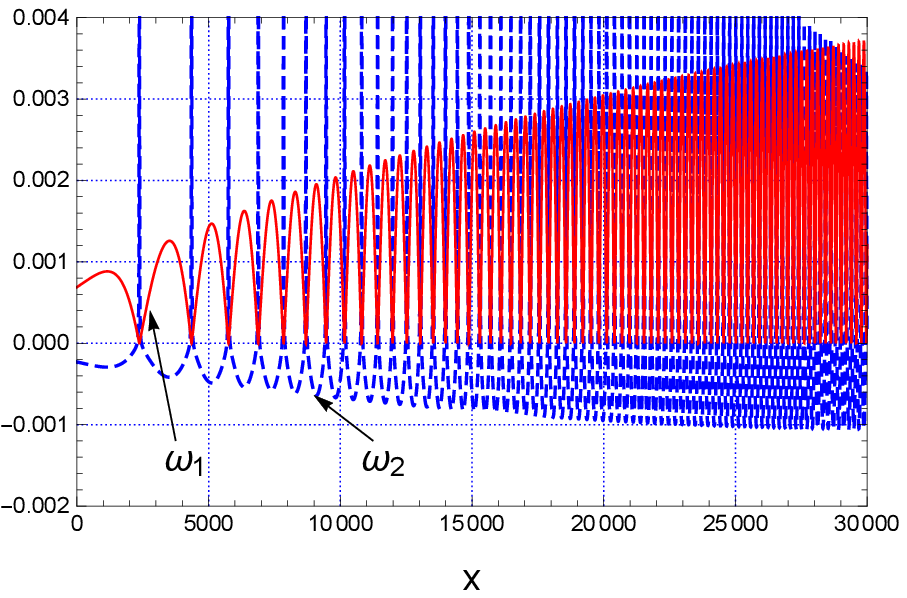}}
\caption{\label{Fig4} The evolution of $\omega_1$ and $\omega_2$ as a function of $x$ in the case of $\xi=-200$.   }
\end{figure}

\begin{figure}
\begin{minipage}[f]{0.5\textwidth}
\includegraphics[width=1\textwidth]{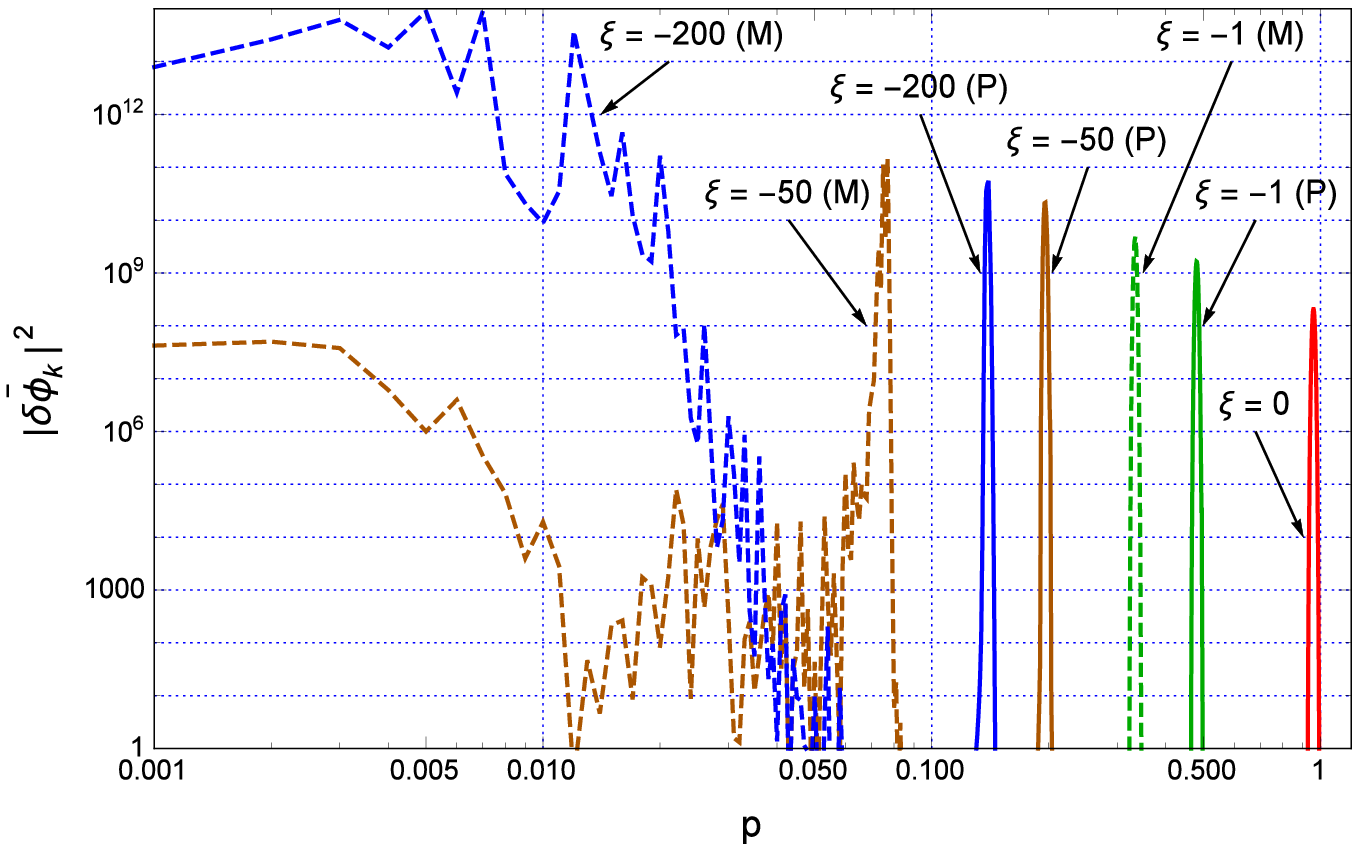}
\end{minipage}
\caption{\label{Fig5} The spectrum of enhanced momentum modes in $\xi=0$, $\xi=-1$, $\xi=-50$ and $\xi=-200$ cases.  M and P represent the metric and Palatini formalisms, respectively. }
\end{figure}

In Figs.~\ref{Fig2}, \ref{Fig3}, and \ref{Fig4},   the evolutions of $\omega_1$ and $\omega_2$ with $x$ are plotted with three different values of $\xi$: $\xi=-1$, $-50$ and $-200$.   We find that the oscillating period of  the scalar field increases with the increase of $|\xi|$, and for a given $\xi$ , it is  shorter in the Palatini formalism than  in the metric formalism, which is a result of the fact, as is shown in Tab.~(\ref{Tab1}), that the values of  $\phi_0(t_i)$ at the beginning of the preheating decrease with the increasing of $|\xi|$ and this decrease is faster in metric formalism than in the Palatini one.  Furthermore, one can see that  in the Palatini formalism  the amplitude of $\omega_1$ is always larger than that of $\omega_2$   although  the decrease of   $\omega_2$  is  slightly lower than that of $\omega_1$  with the increase of $|\xi|$.  While in the metric formalism   with the increase of $|\xi|$   the amplitude of $\omega_2$ will become larger than that of $\omega_1$ at the initial stage of preheating since the significant magnification on $\omega_2$ arises from the non-minimal coupling, and   the domination duration of the $\omega_2$ term becomes longer with a larger $|\xi|$.  Thus,  the structure of resonances is similar to the $\xi=0$ case in the Palatini formalism. However,  the structure changes  with  the increase of $|\xi|$  in the metric formalism .    

To show the changes of  the structure of resonance for different values of $\xi$, we plot Fig.~\ref{Fig5} to give the resonance bands.   With the increase of $|\xi|$, the resonance band in both formalisms moves towards   $k=0$ and it moves faster in the metric formalism. In the Palatini formalism, the resonance band is always narrow. However, in the metric formalism, the resonance band becomes broad for a large $\xi$, and   two effective resonance bands appear when $\xi=-50$. For the strong coupling,  i.e. $\xi\lesssim-100$, the enhanced momentum modes in the metric formalism are mainly low momentum modes.

\begin{figure}[!htb]
                \centering
                \includegraphics[width=0.45\textwidth ]{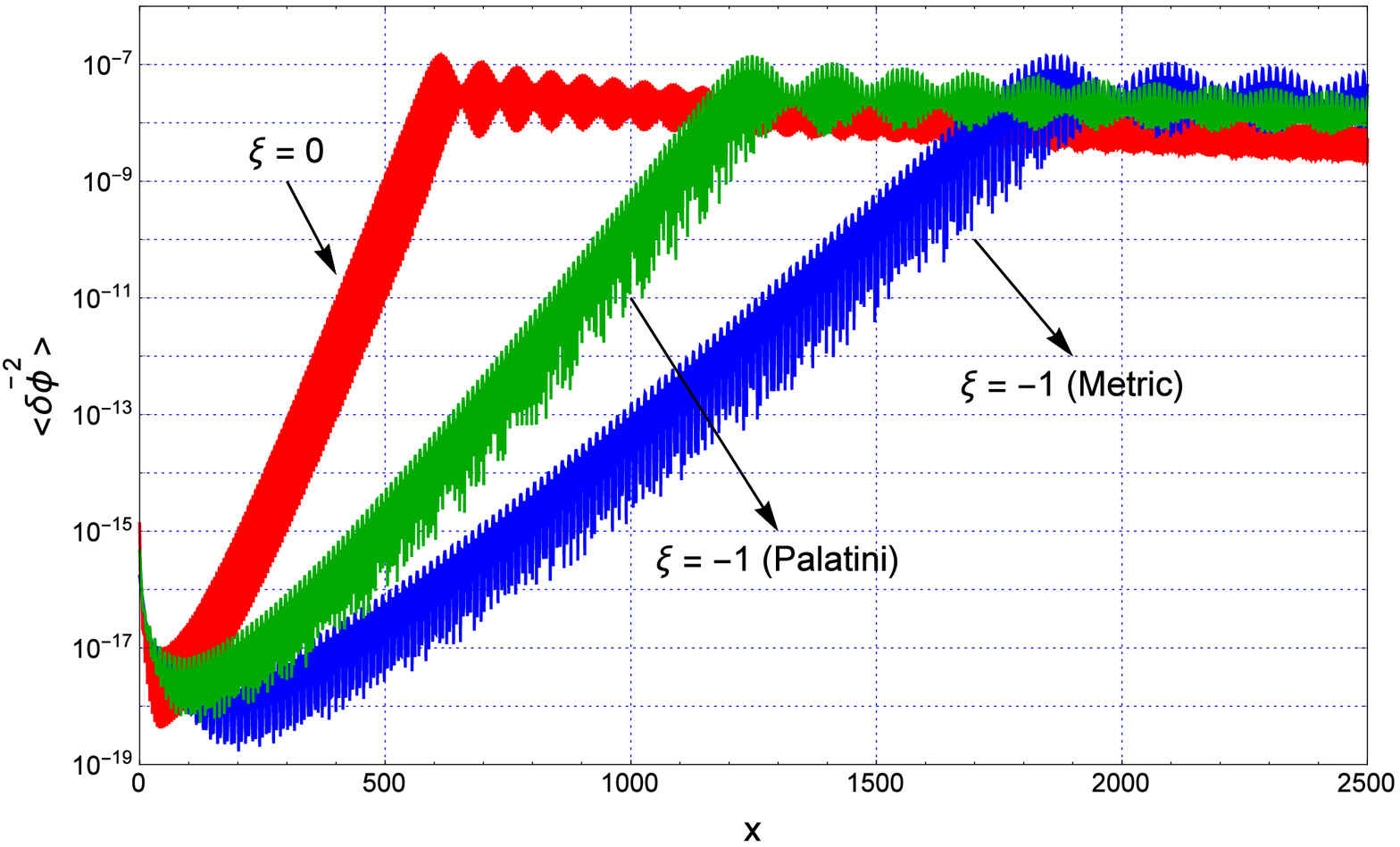}
                  \includegraphics[width=0.45\textwidth ]{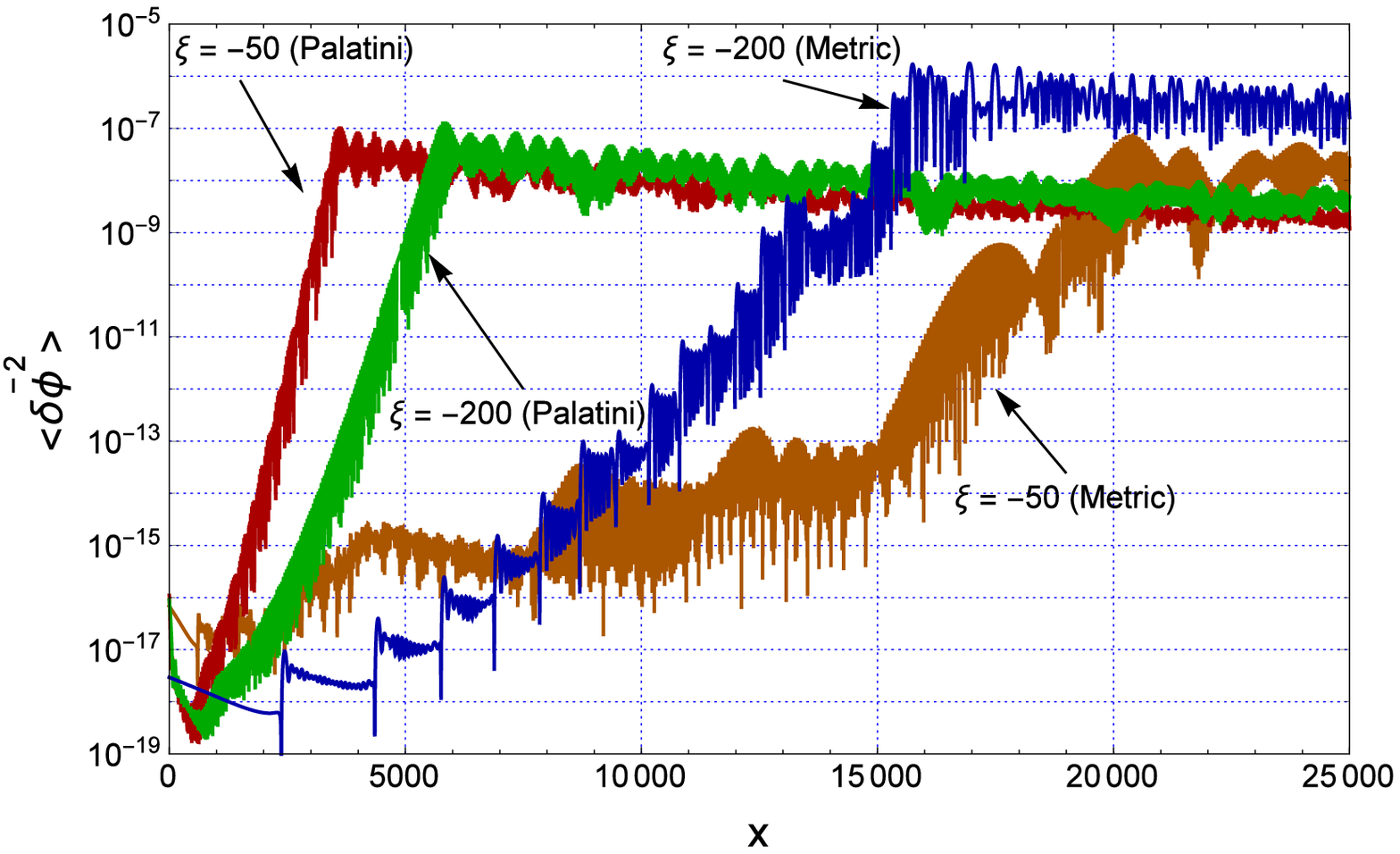}
              \caption{\label{Fig6} The evolution of $\langle\delta\bar\phi^2\rangle$ as a function of $x$ with $\xi=0$ and $\xi=-1$ (left), and $\xi=-50$ and $\xi=-200$ (right).  }
\end{figure}

\begin{figure}
\begin{minipage}[e]{0.5\textwidth}
\includegraphics[width=1\textwidth]{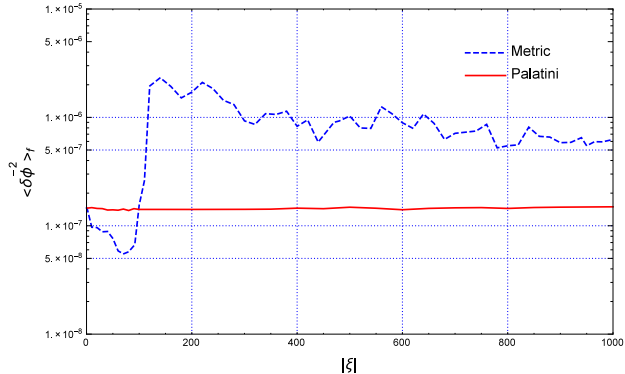}
\end{minipage}
\caption{\label{Fig7} The final value of $\langle\delta\bar\phi^2\rangle$ as a function of $\xi$.  }
\end{figure}

The evolutions of $\langle\delta \bar \phi^2\rangle$  with  $\lambda=10^{-12}$  given by the COBE data are shown in Fig.~\ref{Fig6}. Apparently, the inflaton quanta  do  not grow forever  and the growth will be terminated due to the backreaction,  signaling   the end of the parameter resonance.  We find that in the Palatini formalism, with the increase of $|\xi|$  the growth of the fluctuation is delayed   and the duration of parameter resonance era increases. The final values of the fluctuation ($\langle\delta \bar \phi^2\rangle_f$) are almost the same as that in the $\xi=0$ case, which can also be seen from Tab.~(\ref{Tab2}). For the metric formalism, since $\omega_2$ will play a dominating role during the whole  preheating with the increase of $|\xi|$,   the duration of the parameter resonance era does not increase forever with the increase of $|\xi|$  and the numerical calculation shows that  it has a maximum at about $\xi\simeq -70$, if $-250<\xi<0$. In addition, one can see that  several short plateau appear in the evolution of $\langle\delta\phi^2\rangle$  when $\xi=-50$, which arises from two   effective resonance bands.   Tab.~(\ref{Tab2}) and Fig.~\ref{Fig6} show that the variation of $\langle\delta \bar \phi^2\rangle_f$ is  dramatical in the metric formalism, which is different from that in the Palatini one.    In order to reveal the characteristics of   $\langle\delta\bar \phi^2\rangle_f$  with the increase of $|\xi|$, we plot $\langle\delta\bar\phi^2\rangle_f$ as a function of $|\xi|$ in Fig.~\ref{Fig7}. We find that  in the Palatini formalism the final value of $\langle\delta \bar\phi^2\rangle$ is almost a constant within $-1000\lesssim\xi<0$.  While in the metric formalism $\langle\delta\bar \phi^2\rangle_f$  decreases firstly with the increase of $|\xi|$,  and then reaches a minimum when $\xi$ is about $-70$. After that,  $\langle\delta\bar \phi^2\rangle_f$ increases fast until it reaches its maximum at about $\xi=-120$, and then  decreases slowly.

 Tab.~(\ref{Tab2}) indicates that in the Palatini formalism the oscillation amplitude $\tilde{\phi}_0$ of the inflaton field at the termination time  $x_f$ of the parameter resonance decreases slowly, while in the metric formalism  $\tilde{\phi}_0(x_f)$  does not vary monotonously.  Even so,  in both formalisms  $\langle\delta\bar \phi^2\rangle_f/\tilde{\phi}^2_0(x_f)$ increases with the increase of  $|\xi|$ and it increases faster in the metric formalism.   This indicates that the energy transfer from the inflaton field to $\delta \phi_k$  becomes more efficient with the increase of $|\xi|$ and the efficiency in the Palatini formalism does not increase as fast as that in the metric one.

%%%%%%%%%%%%%%%%%%%%%%%%%%%%%%%%%%%%%%%%%%%%%%%%%%
\section{Conclusion}
\label{sec_co}
%\setcounter{equation}{0}
%%%%%%%%%%%%%%%%%%%%%%%%%%%%%%%%%%%%%%%%%%%%%%%%%%
In this paper, we have studied the   dynamics and preheating in a model  with a non-minimally coupled inflaton field  in the metric and Palatini formalisms.  We find that irrespective of the initial conditions our Universe will evolve into a slow-roll inflationary era in both formalisms. However,  the evolution of the scalar field  in the Palatini formalism deviates from the slow-roll line with  a value of $\phi_0$ much larger than that in the metric formalism.  For the same the  coupling parameter,  the value of the scalar field at the end  of inflation in the Palatini formalism is always larger than that in the metric one, which becomes more and more obvious with the increase of   $|\xi|$.  After inflation, the scalar field rolls into an oscillating phase.  But, the oscillations in different formalisms show distinctive features.

During the preheating,  we find that the inflaton quanta are produced explosively due to the parameter resonance and the growth of inflaton quanta will  be terminated by the backreaction.  In the case of a weak coupling,  the resonance bands  are narrow  in both formalisms. With the increase of $|\xi|$  the resonance bands gradually close to the zero momentum ($k=0$), and the  structure of resonance changes and becomes broader and broader in the metric formalism,  while, it remains to be narrow in the Palatini formalism.     Owing to this effect, when $-1000\lesssim\xi<0$, the final value of $\langle\delta \bar \phi^2\rangle$ in the Palatini formalism does not vary significantly, and is  almost the same as that in the $\xi=0$ case. However, in the metric formalism, when $-70\lesssim\xi<0$,  the final value of the fluctuation  is less than that  in the Palatini formalism and reaches a minimum when $\xi$ is about $-70$. Then,  $\langle\delta \bar \phi^2\rangle_f$ increases fast and becomes larger than that in the Palatini formalism.  It reaches its maximum at about $\xi=-120$, and  then decreases slowly.  In addition, we find that the transfer of the energy from the $\phi_0$ field to the fluctuation is more and more efficient with the increase of $|\xi|$, and  in the metric formalism  the growth of the efficiency of energy transfer is  much faster than that  in the Palatini formalism.
In conclusion, although our Universe dominated by a non-minimally coupled scalar field with a self-interacting potential is sure to evolve into an inflation era and then roll into a preheating phase  in both formalisms, the inflation and preheating processes  show different  characteristics in different formalisms. 

 It is well known that preheating typically results in inhomogeneous configurations of the scalar field and then leads to the production of a stochastic background of gravitational waves~\cite{Kh}, which depends on the resonance structure. So, we  expect that the differences of the non-minimally coupled inflation in different formalisms may imprint on the gravitational waves generated during preheating. We would be in principle  able of differentiating the two different formalisms if such a stochastic background of gravitational waves could be detected  by future high-frequency gravitational wave detectors. The generation of gravitational waves during preheating in the two formalisms is an interesting topic which is currently under our investigation.

\begin{acknowledgments}
This work was supported by the National Natural Science Foundation of China under Grants No. 11775077,  No. 11435006, No.11690034, and  No. 11375092.
\end{acknowledgments}

%%%%%%%%%%%%%%%%%%%%%%%%%%%%%%%%%%%%%%%%%%%%%%%%%%
\appendix
%%%%%%%%%%%%%%%%%%%%%%%%%%%%%%%%%%%%%%%%%%%%%%%%%%

%%%%%%%%%%%%%%%%%%%%%%%%%%%%%%%%%%%%%%%%%%%%%%%%%%
\section{Basic equations in the Palatini formalism}
\label{app:gp}
\setcounter{equation}{0}
%%%%%%%%%%%%%%%%%%%%%%%%%%%%%%%%%%%%%%%%%%%%%%%%%%

In this appendix, we give the basic equation in the Palatini formalism of generalized gravity with the action given in Eq.~(\ref{1}), in which   $F$ is an arbitrary function of   the
scalar field $\phi$. The curvature scalar is derived from $\hat{R}=g^{\mu\nu}\hat{R}_{\mu\nu}$, where the curvature tensor $\hat{R}_{\mu\nu}$ is defined in terms of the torsionless independent connection $\hat{\Gamma}^\alpha_{\beta\gamma}$:
\begin{equation}
\label{A1}
\hat{R}_{\mu\nu} = \hat{\Gamma}^\alpha_{\mu\nu,\alpha} - \hat{\Gamma}^\alpha_{\mu\alpha,\nu} + \hat{\Gamma}^\alpha_{\alpha\lambda}\hat{\Gamma}^\lambda_{\mu\nu} - \hat{\Gamma}^\alpha_{\mu\lambda}\hat{\Gamma}^\lambda_{\alpha\nu}\;,
\end{equation}
where the connection is an independent variable.

Varying the action with respect to the metric tensor $g_{\mu\nu}$  and the connection $\hat{\Gamma}^\alpha_{\beta\gamma}$, respectively, gives
\begin{equation}
\label{A2}
F\hat{R}_{\mu\nu}-\frac{1}{2}F\hat{R} g_{\mu\nu}=\kappa^2T_{\mu\nu}^{(\phi)}\;,
\end{equation}
\begin{equation}
\label{A4}
\hat{\nabla}_\mu(\sqrt{-g}g^{\alpha\beta}F)=0\;,
\end{equation}
where $T^{(\phi)}_{\mu\nu}$ is the energy-momentum tensor of  the scalar field.
Eq.~(\ref{A4}) indicates that a new metric $h_{\mu\nu}$ can be defined by $h_{\mu\nu}=F g_{\mu\nu}$,  which is conformally connected to $g_{\mu\nu}$,  So,  Eq.~(\ref{A4}) becomes
\begin{equation}
\label{A5}
\hat{\nabla}_\mu(\sqrt{-h}h^{\alpha\beta})=0\;.
\end{equation}
The above equation indicates that $\hat\Gamma^\lambda_{\mu\nu}$ is the Levi-Civita connection in regard to the new metric $h_{\mu\nu}$ and has the form
\begin{align}
\label{A6}
\hat{\Gamma}^\lambda_{\mu\nu}&=\frac{1}{2}h^{\lambda\sigma}(h_{\mu\sigma,\nu}+h_{\nu\sigma,\mu}-h_{\mu\nu,\sigma}) \nonumber\\
&=\Gamma^\lambda_{\mu\nu}+\frac{1}{2F}[2\delta^\lambda_{(\mu}\partial_{\nu)}F-g_{\mu\nu}g^{\lambda\sigma}\partial_{\sigma}F]\;.
\end{align}
Here $\Gamma^\lambda_{\mu\nu}$ is the usual Levi-Civita connection defined with $g_{\mu\nu}$.
Thus,  the hatted curvature scalar and Ricci tensor can be  expressed as
\begin{equation}
\label{A7}
\hat{R}_{\mu\nu}=R_{\mu\nu}+\frac{3}{2F^2}(\nabla_\mu F)(\nabla_\nu F)-\frac{1}{F}\nabla_\mu\nabla_\nu F-\frac{1}{2F}g_{\mu\nu}\nabla_\sigma\nabla^\sigma F\;,
\end{equation}
\begin{equation}\label{A8}
\hat{R}=R+\frac{3}{2F^2}(\nabla_\mu F)(\nabla^\mu F)-\frac{3}{F}\nabla_\mu\nabla^\mu F\;,
\end{equation}
 respectively, where $\nabla_\mu$ is the covariant derivative  with respect to $g_{\mu\nu}$.

Using Eq.~(\ref{A7}), we can express the modified Einstein field equations given in Eq.~(\ref{A2}) into the standard form:
\begin{equation}
G_{\mu\nu}=\kappa^2(T_{\mu\nu}^{(\phi)}+T_{\mu\nu}^{(eff)})\;,
\end{equation}
with
\begin{align}
T_{\mu\nu}^{(eff)}=&\frac{1}{\kappa^2}\Bigl[(1-F)R_{\mu\nu}-\frac{3}{2F}(\nabla_\mu F)(\nabla_\nu F)+\nabla_\mu\nabla_\nu F \nonumber \\
&-\frac{1}{2}g_{\mu\nu}\Bigl[(F-1)R-2\nabla_\sigma\nabla^\sigma F+\frac{3}{2F}\nabla_\sigma F\nabla^\sigma F\Bigl]\;.
\end{align}

%%%%%%%%%%%%%%%%%%%%%%%%%%%%%%%%%%%%%%%%%%%%%%%%%%
\section{The action in the Einstein frame}
\label{app:ae}
\setcounter{equation}{0}
%%%%%%%%%%%%%%%%%%%%%%%%%%%%%%%%%%%%%%%%%%%%%%%%%%

We take into account a conformal transformation:
\begin{equation}\label{B1}
\bar{g}_{\mu\nu}=\Omega^2g_{\mu\nu}\;,
\end{equation}
where $\Omega^2\equiv1-\xi\kappa^2\phi^2$. Then we can obtain the following equivalent action:
\begin{equation}\label{B2}
S_E=\int d^4x \sqrt{-\bar{g}}\Bigl[\frac{1}{2\kappa^2}\bar{g}^{\mu\nu}\hat{R}_{\mu\nu}-\frac{1}{2}Q^2(\bar{\nabla}\phi)^2-\bar{V}(\phi)\Bigl]\;,
\end{equation}
where
\begin{equation}\label{B3}
\bar{V}(\phi)\equiv\frac{V(\phi)}{(1-\xi\kappa^2\phi^2)^2}\;,
\end{equation}
and
\begin{equation}\label{B4}
Q^2\equiv\frac{1}{1-\xi\kappa^2\phi^2}+(1-\sigma)\frac{6\xi^2\kappa^2\phi^2}{(1-\xi\kappa^2\phi^2)^2}\;,
\end{equation}

Defining a new scalar field $\psi$:
\begin{equation}\label{B5}
\psi\equiv\int Q(\phi)d\phi\;,
\end{equation}
one achieves the action in the Einstein frame, which has the canonical form:
\begin{equation}\label{B6}
S_E=\int d^4x \sqrt{-\bar{g}}\Bigl[\frac{1}{2\kappa^2}\bar{g}^{\mu\nu}\hat{R}_{\mu\nu}-\frac{1}{2}(\bar{\nabla}\psi)^2-\bar{V}(\psi)\Bigl]\;.
\end{equation}

\end{document}